\documentclass[aps,pre,twocolumn,superscriptaddress,showpacs,10pt]{revtex4-1}
\usepackage[utf8]{inputenc}
\usepackage{t1enc}
\usepackage[T1]{fontenc}
\usepackage{times}
\usepackage{amsmath}
\usepackage{amssymb}
\usepackage{graphicx}
\usepackage{bm}
\usepackage{color}
\usepackage{caption,subcaption}
\usepackage{hyperref}

\newcommand{\ket}[1]{\left|#1\right>}
\newcommand{\bra}[1]{\left<#1\right|}

\begin{document}

\title{Transport properties of continuous-time quantum walks on Sierpinski fractals}

\author{Zolt\'an Dar\'azs}
\thanks{These authors contributed equally to this work.}
\affiliation{WIGNER RCP, SZFKI, Konkoly-Thege Mikl\'os \'ut 29-33, H-1121 Budapest, Hungary}
\affiliation{E\"otv\"os University, P\'azm\'any P\'eter s\'et\'any 1/A, H-1117 Budapest, Hungary}
\author{Anastasiia Anishchenko}
\thanks{These authors contributed equally to this work.}
\affiliation{
Physikalisches Institut, Universit\"at Freiburg,
Hermann-Herder-Stra{\ss}e 3, 79104 Freiburg, Germany}
\author{Tam\'as Kiss}
\affiliation{WIGNER RCP, SZFKI, Konkoly-Thege Mikl\'os \'ut 29-33, H-1121 Budapest, Hungary}
\author{Alexander Blumen}
\author{Oliver M\"ulken}
\affiliation{
Physikalisches Institut, Universit\"at Freiburg,
Hermann-Herder-Stra{\ss}e 3, 79104 Freiburg, Germany}


\date{\today}

\begin{abstract}

We model quantum transport, described by continuous-time quantum walks (CTQW), on deterministic Sierpinski fractals, 
differentiating between Sierpinski gaskets and Sierpinski carpets, along with their dual structures. The
transport efficiencies are defined in terms of the exact and the average return probabilities, as well as by the mean survival
probability when absorbing traps are present. In the case of gaskets, localization can be identified already for small networks
(generations). For carpets, our numerical results indicate a trend towards localization, but only for relatively large structures.
The comparison of gaskets and carpets further implies that, distinct from the corresponding classical continuous-time
random walk, the spectral dimension does not fully determine the evolution of the CTQW.

\end{abstract}

\maketitle

\section{Introduction}

Networks are sets of connected nodes \cite{havlin2010complex,albert2002statistical}, and their static and the dynamic properties
are of much interest. Applications range from, say, polymer science \cite{doi1988theory}, over traffic and power grid studies
\cite{albert2004structural}, up to social networks \cite{yu2003searching}. A special class of networks are deterministic fractals
which as such can be built iteratively. We remark that for them sometimes analytic results can be obtained, see, e.g.,
Refs.~\cite{Agliari2008, blumen2002multifractal, cosenza1992coupled}.

Now, the classical dynamics of random walks (RW) over networks has been extensively investigated in the last decades
\cite{van2011stochastic,weiss1994aspects}. This effort has led to a very detailed understanding of the influence of the network's
topology on RW. When the efficiency of transport is concerned, the question whether the RW is recurrent or transient boils down 
to determining the probability of the RW to return to its origin, which is also related to the P\'olya number
\cite{polya1921aufgabe}. Moreover, the global properties of the RW can also be captured by introducing the local probability 
decay channels and calculating the averaged decay time of the excitation, known as the averaged mean first passage time (MFPT)
\cite{van2011stochastic,condamin2007first}. For simple undirected networks the transfer matrix of the continuous-time random walk
(CTRW) is given by the connectivity matrix of the network \cite{mulken2011continuous}. Many networks show scaling behavior for 
the lower part of the spectrum of the connectivity matrix, with an exponent $d_s$ which is called {\it spectral dimension}
\cite{alexander1981excitation}. As it turns out, $d_s$ determines many of the dynamical properties of the network, e.g., the
return to the origin or the MFPT.

For the quantum mechanical aspects of transport on networks, we choose as a model the continuous-time quantum walk
(CTQW), which is related to the classical  CTRW \cite{mulken2011continuous}. In this way, the Hamiltonian is determined
by the connectivity of the network. Therefore, by analyzing the connectivity matrix, we obtain results for both, CTRW
and CTQW. While in recent years CTQW over several types of networks have been analyzed \cite{mulken2011continuous},
there is no unambiguous classification according to, say, the spectral dimension. In many aspects, the quantum dynamics
is much richer (i.e., more complex) than the classical CTRW counterpart, since it also involves the wave properties of
the moving object. In several cases of tree-like networks, such as stars \cite{Muelken2006,Anishchenko_QIP_Stargraph} or
dendrimers \cite{muelken2006coherent}, it has been shown that the (average) quantum mechanical transport  efficiency,
defined by the return to the origin, is rather low compared to structures which are translationally invariant.  Quantum
walks are interesting models also from the point of view of quantum information processing \cite{venegas2012quantum}.
Search via quantum walks on fractal graphs has been considered in  Refs.~\cite{Agliari2008, agliari2010quantum,
patel2012search}.

A similar mathematical model arises for condensed matter systems, in which one considers a particle moving on an underlying fractal
lattice (a Sierpinski gasket); here the solution of Schr\"odinger's equation has been studied within the tight-binding
approximation \cite{domany1983solutions,wang1995localization}. For several fractals considered, the dynamics has been shown to be
subject to localization effects, similar to the classical waves in fractal waveguides \cite{li2000photonic}. From an experimental
point of view, recent years have seen a growing number of possible implementations of CTQW, for example, using interference
effects of light. Those experiments range from photonic waveguides \cite{rechtsman2013photonic} to fiber-loops
\cite{schreiber20122d}.

In this paper we study quantum transport over fractal networks, namely over Sierpinski gaskets (SG) and their dual
structures (DSC) as well as over Sierpinski carpets (SC) and their dual structures (DSC). In the case of the SG and
of their duals we find clear signatures of localization around the initial starting node, indicating recurrent behavior.
We seek to answer  the question whether the spectral dimension $d_s$ of the graph determines the transport properties
for  CTQW. Given the great experimental control over, say, coupling rates and decoherence, we believe that our results
for fractal structures can also be experimentally realized, say, through photonic waveguides.

The paper is organized as follows, Sec.~\ref{methods} gives an overview over the quantities we use to determine the
performance of CTQW over networks. In Sec.~\ref{systems} we outline the deterministic construction rules of the
SG and SC and their dual transformations, along with their spectral properties. These systems are
then analyzed in detail in Secs.~\ref{DSG}-\ref{SC}. We close with a summary of results in Sec.~\ref{summary}.

\section{Methods}\label{methods}
 
We model the quantum dynamics of an excitation over a given fractal network by CTQW and compare this to its classical
counterpart, the CTRW, over the same network. A network is determined by a set of $N$ nodes and a set of bonds. With
each of the nodes we associate a state $|k\rangle$ corresponding to an excitation localized at node $k$. For both, CTQW
and CTRW, the dynamics is determined by the network's connectivity, i.e., by its connectivity matrix $\mathbf A$. The
off-diagonal elements of $\mathbf{A}$ are $A_{kj}=-1$  if the nodes $k$ and $j$ are connected by a single bond and are
$A_{kj}=0$ otherwise; the diagonal elements are $A_{kk}=f_k$, where $f_k$ is the functionality of node $k$, i.e., the
number of nodes connected to $k$ through a single bond. The matrix $\mathbf A$ is real and symmetric and has only real
and non-negative eigenvalues. For networks without disjoint parts all eigenvalues  are positive except one,
$E_{min}=0$.

Now, we take for CTRW the transfer matrix $\mathbf T = - \mathbf A$ and for CTQW the Hamiltonian $\mathbf H = \mathbf A$
(i.e. in the following we set $\hbar=1$ and normalize the transfer capacity of each bond to unity), see also
\cite{farhi1998quantum,mulken2011continuous}, such that the transition probabilities read $p_{k,j}(t) = \langle k |
\exp(\mathbf T t) | j \rangle$ and $\pi_{k,j}(t) = |\langle k | \exp(-i \mathbf H t) | j \rangle |^2$, respectively. By
diagonalizing $\mathbf A$ we obtain the eigenvalues $E_n$ and the eigenstates $|\Phi_n\rangle$ (with $n=1,\dots,N$) of
$\mathbf A$, resulting in
\begin{gather}
p_{k,j}(t) = \sum_{n=1}^N \exp(- E_n t) \langle k |\Phi_n \rangle \langle \Phi_n | j \rangle 
\label{eq:ret-prob-cl}
\end{gather}
for CTRW and
\begin{gather}
\qquad \pi_{k,j}(t) = \left | \sum_{n=1}^N \exp(-i E_n t)  \langle k | \Phi_n \rangle \langle \Phi_n | j \rangle \right|^2
\label{eq:ret-prob-q}
\end{gather}
for CTQW. In principle all quantities of interest can be calculated on the basis of the transition probabilities. 
In order to quantify the efficiency of the transport, we will focus on three quantities: the exact return probability and the
related P\'olya number, the average return probability, and the mean survival probability. 

\subsection{P\'olya number}

The so-called P\'olya number allows to assess the {\sl local} transport properties. In classical systems, the definition
of the recurrence  is straightforward: it characterizes the event that the walker returns to its initial position. For
quantum walks one can imagine different definitions depending on the envisaged measurement procedure
\cite{Kollar_Scripta_FullRevivals, Kollar_PRA_ThreeState,Stefanak_PRL_DTQWPolya, Stefanak_NJP_RecurrenceBiased,
Zhang2011Polya, Wan2012Polya, grunbaum2013recurrence}.

Ref.~\cite{Darazs2010polya} suggests a possible quantum definition for the P\'olya number, which is directly related to the
return probability to the initial node ($\ket{\psi(0)}=\ket{1}$):
\begin{gather}
	\pi_{1,1}(t) = \left | \bra{1}\exp(-i\mathbf{H}t)\ket{1}  \right |^2 \, .
\end{gather}
The formal definition of the P\'olya number reads
\begin{gather}
\mathcal{P} = 1-\prod_{i=1}^{\infty}[1-\pi_{1,1}(t_i)] \, ,
\end{gather}
where the set $\{t_i, i=1,\dots \infty\}$ is an infinite time series which can be chosen regularly or be determined by
some random process. It can be shown that its value depends on the convergence speed of $\pi_{1,1}(t)$ to zero: if
$\pi_{1,1}(t)$ converges faster than $t^{-1}$ then the CTQW is transient, otherwise it is recurrent
\cite{Darazs2010polya}.

For a finite network of $N$ sites the probability that we find the walker at the origin can be written as a finite sum of cosine 
functions,
\begin{eqnarray}
 \hspace{-5mm} \pi_{1,1}(t)&=&\left | \sum_{n=1}^{N} \langle 1 | e^{-iE_nt} | \Phi_n \rangle \langle \Phi_n | 1 \rangle \right |^2 
 \nonumber \\
&=&\sum_{n,m=1}^{N} |\langle 1 |\Phi_n \rangle |^2 |\langle 1 |\Phi_m \rangle |^2 \cos \left [(E_m-E_n) \:\! t \:\! \right  ] \, .
\label{eq:exact-return-prob}
\end{eqnarray}
A finite sum of cosine functions cannot be a decaying function of time and thus for any finite system the P\'olya number
equals one, meaning that the  walk is recurrent. On the other hand, in an infinite network ($N\to\infty$),
$\pi_{1,1}(t)$ might tend to zero in the $ t \to \infty$ limit. If the return probability has the asymptotic form
$\pi_{1,1}(t) \sim f(t) \cdot t^{-\delta}$ where $f(t)$ is a periodic or an almost periodic analytical function, then,
with regular and Poissonian sampling, the walk is recurrent if $\delta\leq 1$, and it is transient if the envelope
decays faster ($\delta > 1$) \cite{Darazs2010polya}. For CTRW on the fractals considered in the following, the decay of
the probability $p_{1,1}(t)$ is slower than $t^{-1}$, which can be seen from the fact that on a fractal $p_{1,1}(t)$
scales as $t^{-d_s/2}$ and the fractals considered in this paper have spectral dimension $d_s < 2$
\cite{Gasket_spectraldimension,Barlow_Resistance}.

\subsection{Average return probability}

As a {\sl global} efficiency measure, the average return probability is defined as the probability to remain or return to the
initial node $j$,  averaged over all nodes:
\begin{eqnarray}
\label{eq:average-return-prob-cl}
\overline p (t) &\equiv& \frac{1}{N} \sum_{j=1}^N p_{j,j}(t) 
\\
\mbox{and} \qquad \overline \pi (t) &\equiv& \frac{1}{N} \sum_{j=1}^N \pi_{j,j}(t) .
\label{eq:average-return-prob-q}
\end{eqnarray}
While $\overline p (t)$ only depends on the eigenvalues, $\overline \pi (t)$  also depends on the eigenstates. However, by using the 
Cauchy-Schwarz inequality a lower  bound, independent of the eigenstates, has been introduced in \cite{Muelken2006}:
\begin{equation}
\overline \pi (t) = \frac{1}{N} \sum_{j=1}^N \pi_{j,j}(t) \geq \Big| \frac{1}{N} \sum_{j=1}^N \alpha_{j,j}(t) \Big|^2 \equiv \Big|\
\overline\alpha(t)\Big|^2.
\label{eq:lower_bound_q}
\end{equation}
In Eq.~(\ref{eq:lower_bound_q}) $\alpha_{j,j}(t) = \langle j | \exp(-i \mathbf H t) | j \rangle$ is the transition
amplitude  between two nodes. In the following we will compare $\overline p (t)$  with $\big|\overline\alpha(t)\big|^2$
and express both quantities in terms of the (discrete) density of states (DOS)
\begin{equation}
\tilde {\rho}(E) = \frac{1}{N} \sum_{n=1}^{N} \delta(E - E_n) \, .
\label{eq:DOS}
\end{equation}
 Here $\delta(E - E_n)$ is the Dirac delta-function. Integrating $\tilde{\rho}(E)$ in a very small neighborhood of an
  eigenvalue, say, $E_m$, gives
 \begin{equation}
\lim_{\varepsilon \to 0^+}\int_{E_m-\varepsilon}^{E_m+\varepsilon}\tilde{\rho}(E)\, \mathrm{d}E=D(E_m)/N\equiv\rho(E_m) \, .
 \label{eq:intDOS}
\end{equation}
where $D(E_m)$ is the degeneracy of $E_m$ and we introduced $\rho(E)$. This yields
\begin{eqnarray}
\nonumber
 \overline p (t) &=&\sum_{\{E_m\}} \rho(E_m)\exp(- E_m t) \\
&=&\int_{-\infty}^{\infty} \tilde{\rho}(E)\exp(-Et) \, \mathrm{d}E
\end{eqnarray}
and
\begin{eqnarray}
\nonumber
\big|\overline\alpha(t)\big|^2 &=& \Big|\sum_{\{E_m\}}\rho(E_m)\exp(- i E_m t)\Big|^2  \\
 &=& \Big|\int_{-\infty}^{\infty} \tilde{\rho}(E)\exp(-iEt) \, \mathrm{d}E \, \Big|^2 \,,
\end{eqnarray}
where the sums run over the set $\{E_m\}$ of {\sl distinct} eigenvalues.

Now, if both $\overline p (t)$ and $\big|\overline\alpha(t)\big|^2$ decay very quickly in time, the average
probability to find the excitation at any node but the initial node increases quickly. Then we call the transport over
the network efficient, because {\bf (}on average{\bf )} the excitation will efficiently explore parts of the network
away from the initial node. In contrast, if these quantities decay  very slowly, we regard the transport as being inefficient.

For CTRW and not too short times, $\overline p (t)$ is dominated by the small eigenvalues. For fractals, the DOS typically scales with
the so-called spectral dimension $d_s$ \cite{alexander1981excitation}, i.e., $\tilde{\rho}(E) \sim E^{d_s/2-1}$. Then, one finds in an
intermediate time range, before the equilibrium value is reached, that $\overline p (t) \sim t^{-d_s/2}$. However, for  CTQW such a
simple analysis does not hold due to the coherent evolution. Instead, highly degenerate eigenvalues dominate
$\big|\overline\alpha(t)\big|^2$, see Ref.~\cite{Muelken2007Inefficient,domany1983solutions}. In the case that one has a single highly
degenerate eigenvalue $E_m$, the lower bound of the average return probability can be approximated by \cite{Anishchenko_QIP_Stargraph}
\begin{gather}
\nonumber
|\bar{\alpha}(t)|^2 \approx \tilde{\rho}^2(E_m) + \tilde{\rho}(E_m) \lim_{\varepsilon\to 0^+}\Bigg [\int_{-\infty}^{E_m-\varepsilon}
 \tilde{\rho}(E) \\ \cos \big [ (E-
E_m)t \big ] \mathrm{d}E + \int_{E_m+\varepsilon}^{\infty} \tilde{\rho}(E) \cos \big [ (E-E_m)t \big ] 
\mathrm{d}E \Bigg ]\, .
\label{eq:DOS_localization}
\end{gather}
If there is at least one eigenvalue for which $\tilde{\rho}(E_m)$ is $\mathcal{O}(1)$, then the average transition
amplitude does not tend to zero. Then the long time average $\overline\chi_{lb}$ of the transition probability also
allows to quantify the global performance of CTQW through \cite{Anishchenko_QIP_Stargraph}
\begin{gather}
\label{eq:DOS-lta}
	\overline\chi_{lb}= \lim_{T\to\infty} \frac{1}{T} \int_0^T \! |\bar{\alpha}(t)|^2 \, \mathrm{d}t =
	 \sum_{\{E_m\}} \left [ \tilde{\rho}(E_m) \right ]^2 \, .
\end{gather}

\subsection{Mean survival probability}

In order to corroborate our findings for the average return probabilities, we define another (global) transport
efficiency measure which is based  on the mean survival probability, see also \cite{Muelken2007} for CTQWs and 
\cite{Yutaka_Survival_2011} for discrete time quantum walks. Here, the original
network is augmented by local decay channels which act as traps for the walker. These traps are localized at a set 
$\cal M$ of nodes $m$ of the original network. For this the total number of nodes of the system is not changed, but the
transfer matrix $\mathbf T$ as well as the Hamiltonian $\mathbf H$ get augmented by additional terms, such that the new
(effective) matrices read $\mathbf T_{\rm eff} \equiv \mathbf T - \mathbf\Gamma$   and $\mathbf H_{\rm eff} \equiv
\mathbf H - i \mathbf\Gamma$, respectively, where the trapping matrix is diagonal, namely  $\mathbf \Gamma = \Gamma
\sum_{m\in\cal M} | m \rangle\langle m|$ with a trapping rate $\Gamma$ which we set equal for all traps. We note that
such an effective Hamiltonian can be obtained within the framework of quantum master equations of Lindblad type, where
the network is only coupled to the environment at the trap nodes, see Ref.~\cite{Schijven2012Modeling}. For CTRW, such
traps will still lead to a real symmetric transfer matrix, but now with only positive eigenvalues
\cite{van2011stochastic}. For CTQW, the new Hamiltonian $\mathbf H_{\rm eff}$ becomes non-Hermitian. Such Hamiltonians
can have complex eigenvalues $E_n = \epsilon_n - i\gamma_n$ with a real part $\epsilon_n$ and an imaginary part
$\gamma_n$. As has been shown in \cite{Muelken2007}, by averaging the transition probabilities over all possible initial
and final nodes one obtains the mean survival probability for CTQW as a function solely of the $\gamma_n$,
\begin{equation}
\Pi(t) \equiv \frac{1}{N} \sum_{j,k=1}^N \pi_{k,j}(t)= \frac{1}{N} \sum_{n=1}^N \exp(-2 \gamma_n t) \, .
\label{eq:surv-prob-ctqw}
\end{equation}
Note the slightly different definition of $\Pi(t)$ compared to the one in Ref.~\cite{Muelken2007}. Here we do not
exclude the trap nodes from the sum, thus Eq.~(\ref{eq:surv-prob-ctqw}) becomes exact. For CTRW a similar approach with
the new transfer matrix $\mathbf T_{\rm eff}$ yields \cite{Muelken2008}
\begin{equation}
P(t) \equiv \frac{1}{N} \sum_{j,k=1}^N p_{k,j}(t) = \frac{1}{N} \sum_{n=1}^N \exp(-\lambda_n t) \Big|\sum_{j=1}^N \langle j |
\Psi_n\rangle\Big|^2,
\label{eq:surv-prob-ctrw}
\end{equation}
where $\lambda_n$ and $\ket{\Psi_n}$ are the eigenvalues and eigenstates of $\mathbf T_{\rm eff}$,
respectively. Thus, $P(t)$ will eventually decrease to zero and the asymptotical behavior will be dominated by the
smallest eigenvalue. Now, if $\Pi(t)$ and $P(t)$ decrease quickly we also call the transport
efficient (on average) because then an initial excitation will reach the trap rather quickly.

For CTQW one can relate the $\gamma_n$ to the eigenstates of the original Hamiltonian $\mathbf H$ within a
 (non-degenerate) perturbative treatment, 
$\gamma_n = \Gamma \sum_{m\in\cal M} \big| \langle m | \Phi_n \rangle \big|^2$
\cite{mulken2011continuous}. Thus, the imaginary parts $\gamma_n$ are determined by the overlap of the eigenstates
$|\Phi_n\rangle$ of $\mathbf H$ with the locations of the  traps. This implies that for  localized eigenstates this overlap can be
zero, such that for some $n$ the imaginary parts vanish, $\gamma_n=0$. This yields a mean survival probability  which
does not decay to zero but which reaches the asymptotic value
\begin{equation}
\Pi_\infty \equiv \lim_{t\to\infty} \Pi(t) = \frac{N_0}{N} \, ,
\label{eq:pi-limit}
\end{equation}
where $N_0$ is the number of eigenstates for which the $\gamma_n$ vanish. For the SG it has been shown
that such eigenstates exist,  which in fact gives rise to localization effects \cite{domany1983solutions}.

We have now defined the asymptotic quantity $\Pi_\infty$ which allows us to assess the transport
properties of CTQW by calculating the probability that the walker will stay forever in the network.

\section{The systems under study}\label{systems}

\begin{center}
\begin{figure}[ht!]
	\begin{subfigure}[(a)]{0.22\textwidth}
                \includegraphics[height=\textwidth]{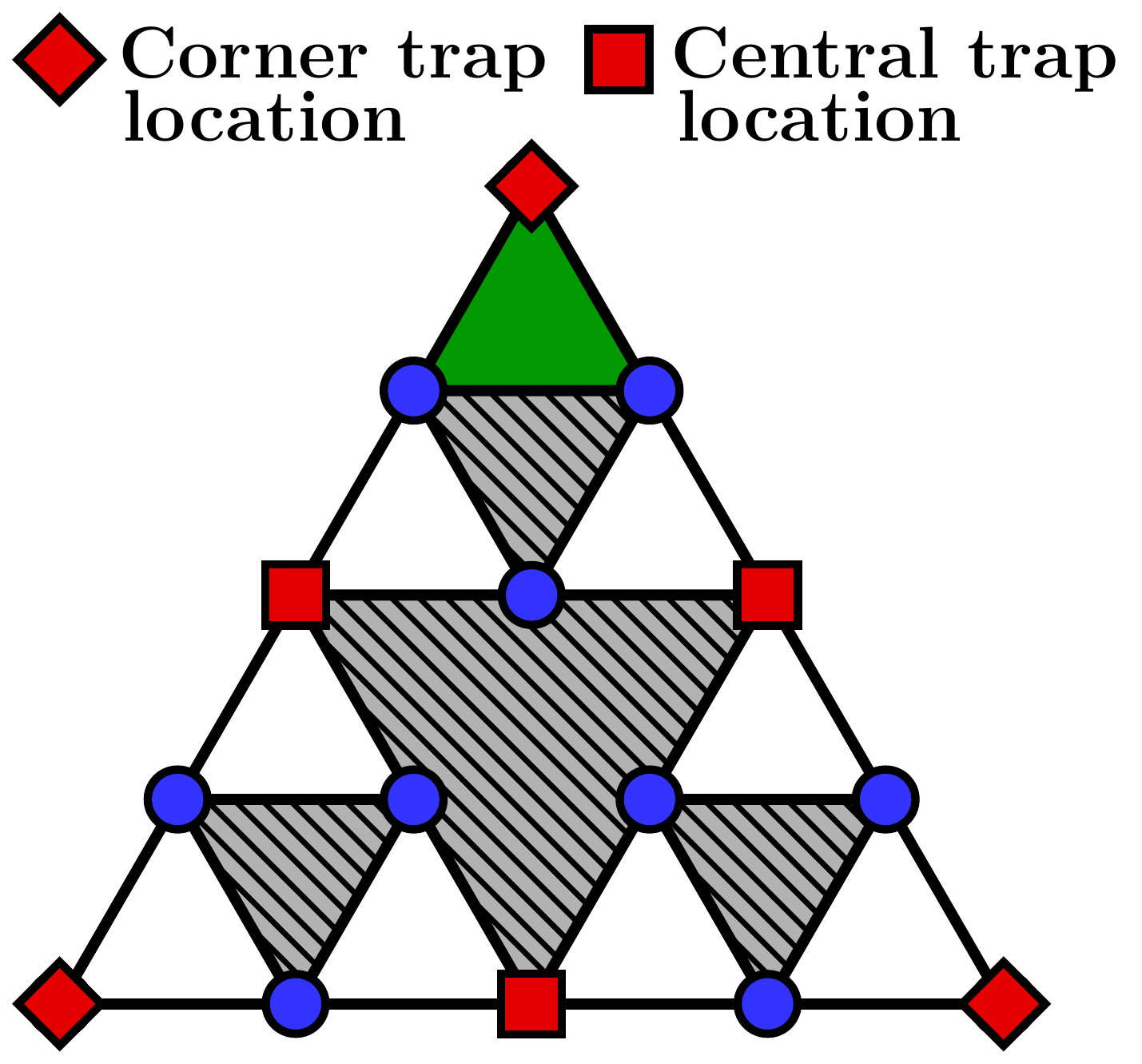}
                \caption{SG}
                \label{fig:SG}
    \end{subfigure}
    \hspace{2mm}
    \begin{subfigure}[(b)]{0.2\textwidth}
                \includegraphics[height=\textwidth]{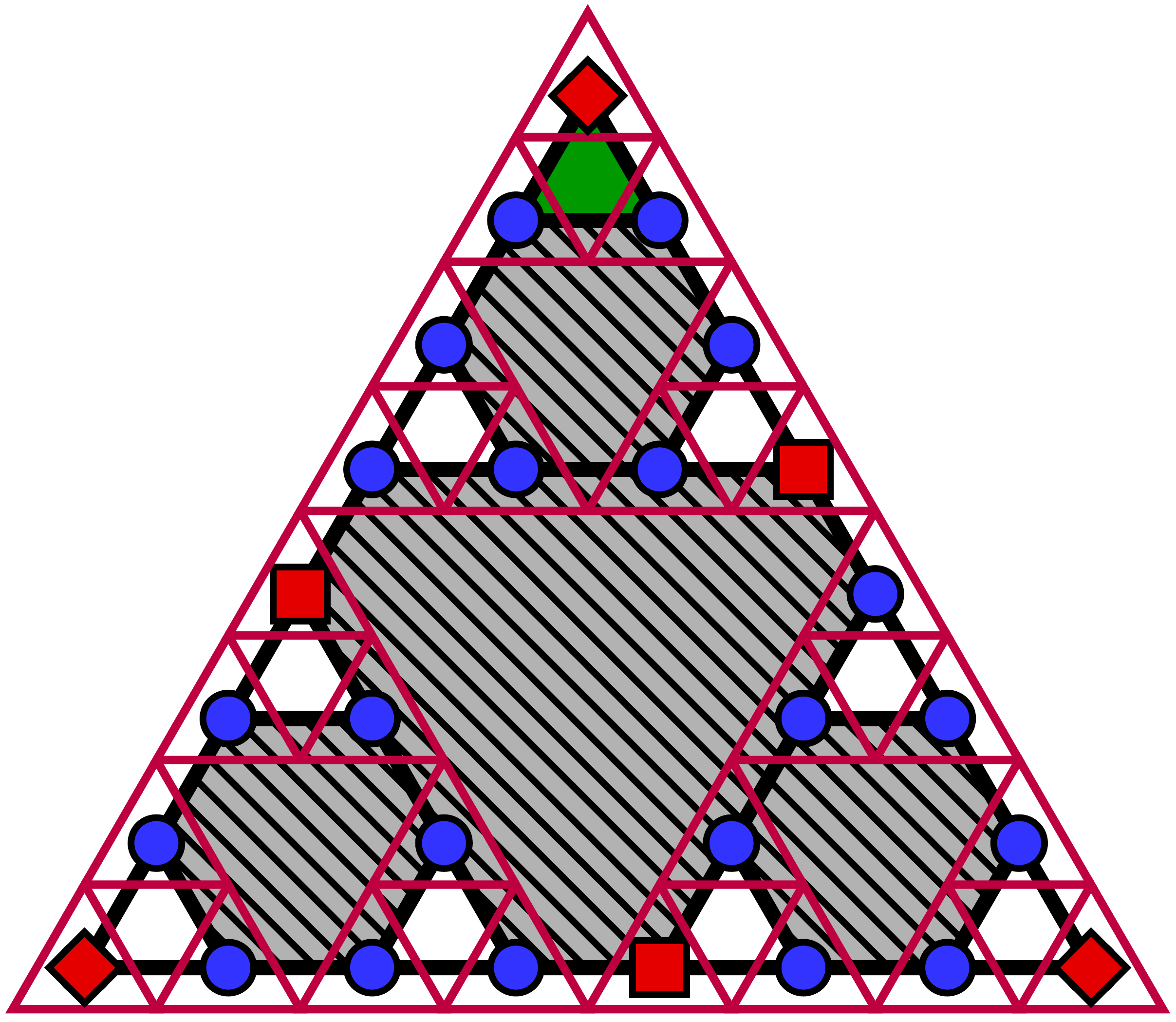}
                \caption{DSG}
                \label{fig:DSG}
    \end{subfigure}
    \begin{subfigure}[(c)]{0.235\textwidth}
                \includegraphics[height=\textwidth]{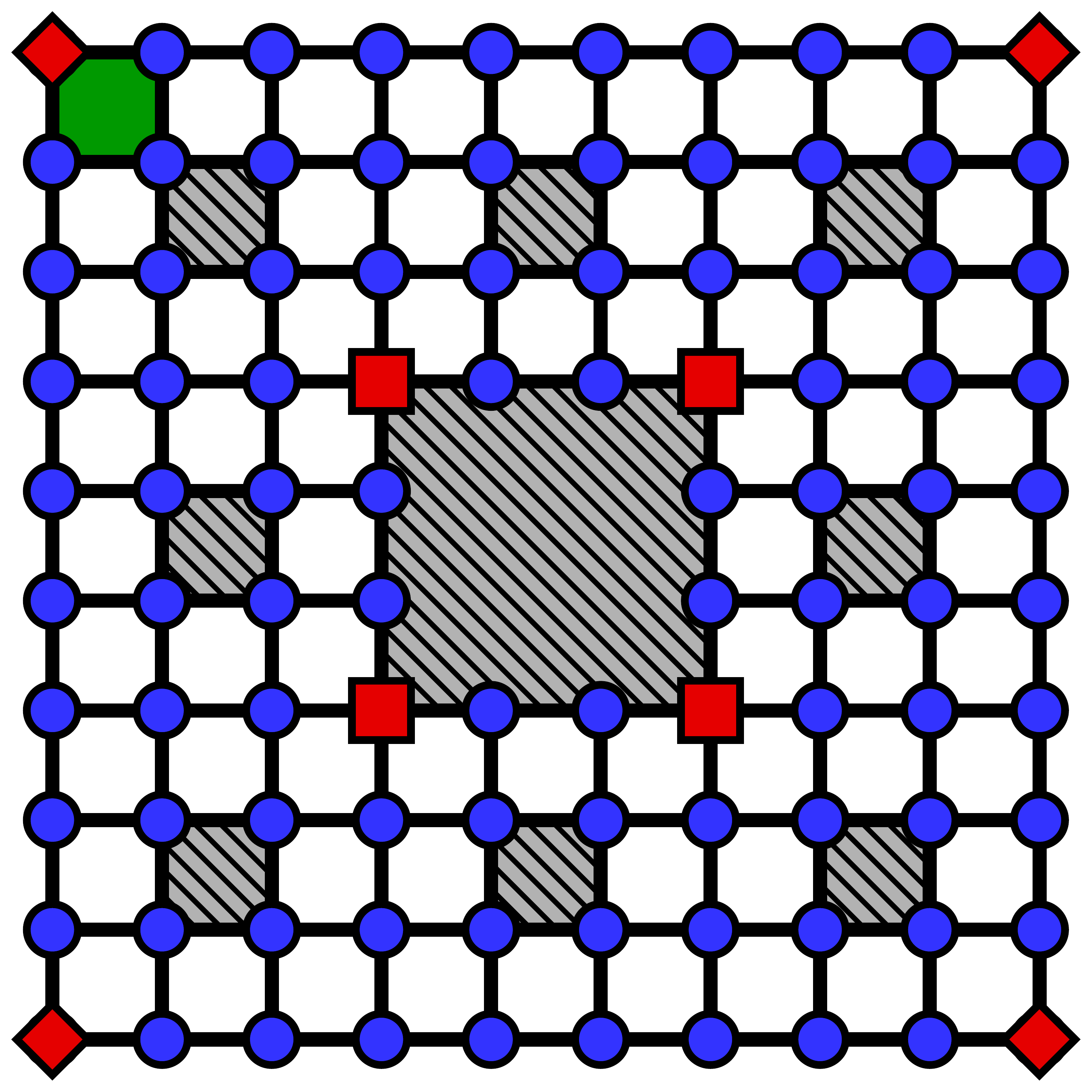}
                \caption{SC}
                \label{fig:SC}
    \end{subfigure}
    \hfill
    \begin{subfigure}[(d)]{0.235\textwidth}
                \includegraphics[height=\textwidth]{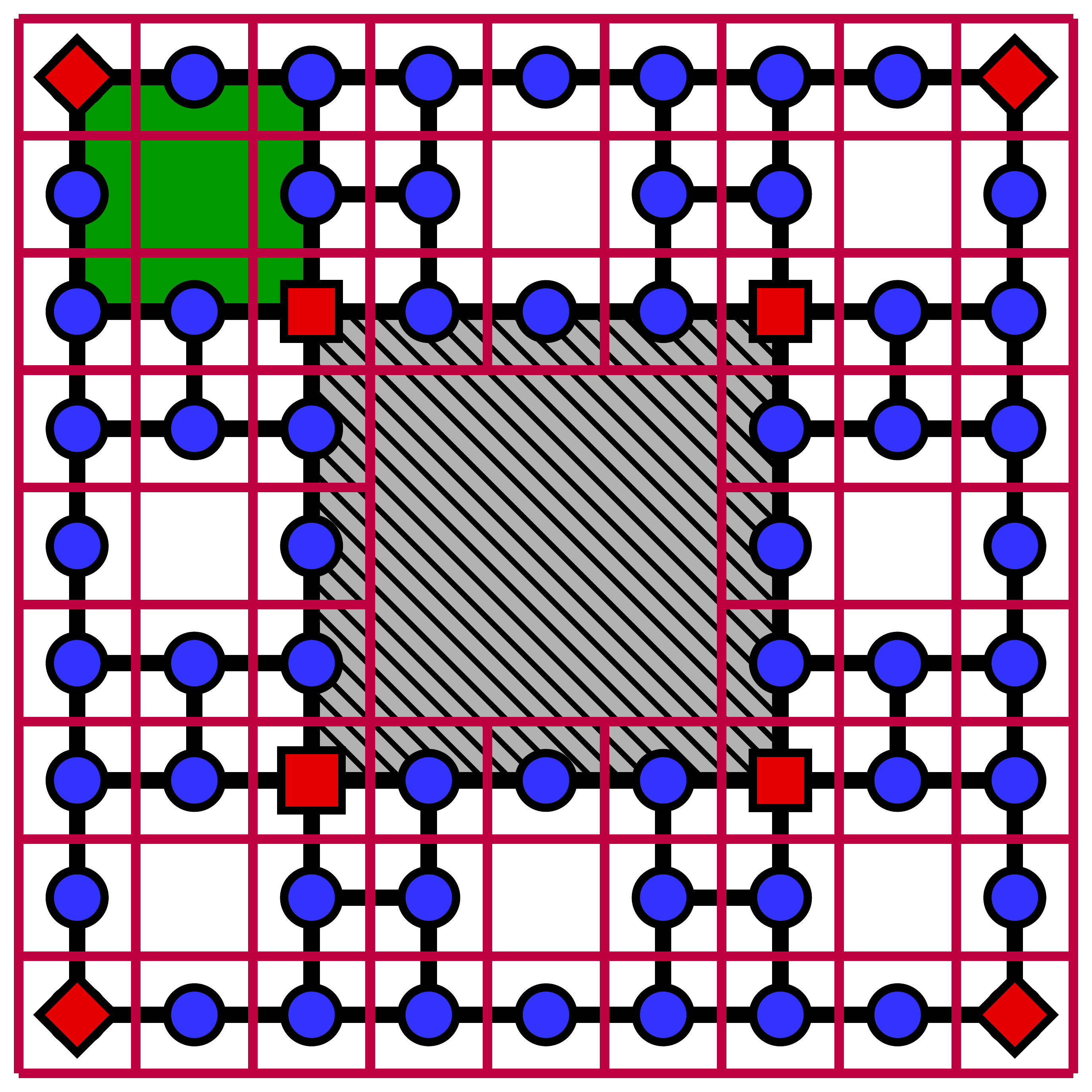}
                \caption{DSC}
                \label{fig:DSC}
    \end{subfigure} 
    \caption{(Color online) The graphs under study. The graphs are at third generation ($g=3$), except the DSC for
		which $g=2$. We denoted the $g=1$ graphs with green, and the holes with a gray (striped) background. Traps are put either
		at the positions indicated by the small diamonds or at the positions indicated by the small squares.}
    \label{fig:fractals}
\end{figure}
\end{center}

We now discuss the systems under study and their topological properties. We consider two groups of Sierpinski fractals,
namely, gaskets and carpets, along with their dual transformations. These fractals are built in an iterative manner: In
order to construct the SG, one starts from a triangle of three nodes. In the next step two additional
triangles  are attached to the corner nodes by merging them, so that they form a bigger self-similar triangle.  The
procedure is then iterated, see Fig.~\ref{fig:fractals}(a) for a gasket at generation $g=3$.  A similar idea is
used for creating SC, where instead of triangles the central building blocks are squares, see also
Fig.~\ref{fig:fractals}(c). At generation  $g$ the total number of nodes of the SG is $N_{\rm SG}=(3^g+3)/2$ and of the
SC is $N_{\rm SC}=\frac{11}{70}\cdot8^g+\frac{8}{15}\cdot3^g+\frac{8}{7}$,  so that at the same (large) $g$ the carpet
has much more nodes than the gasket.

The dual networks of the Sierpinski fractals are easily obtained by the following procedure:  In the original structure
one replaces each of the smallest building blocks (triangles for gaskets and squares for carpets)  by a node and
connects then the  nodes which belong to building blocks sharing a node (for carpets we only allow connections in the
horizontal and in the vertical direction but not diagonally), see also Figs.~\ref{fig:fractals}(b)  and
\ref{fig:fractals}(d), which illustrate the procedure by also showing the underlying lattices of the SG and SC,
respectively. The number of nodes of the DSG of generation $g$ is $N=3^g$ and of the DSC of generation $g$ is $N=8^g$.

Based on real space renormalization arguments, one can show that a structure and its dual have the same fractal $d_f$
and spectral $d_s$ dimensions.  For the SG and the DSG,  the corresponding values are
$d_f=\rm{ln}(3)/\rm{ln}(2)\approx 1.5849\dots$ and $d_s=2\rm{ln}(3)/\rm{ln}(5)\approx 1.3652\dots$, see
Ref.~\cite{Agliari2008}. For the SC and the DSC, one has $d_f=\rm{ln} (8)/\rm{ln}(3)\approx1.8928\dots$ and
$d_s\approx 1.805$ \cite{Barlow_Resistance}.

For our calculations of the average return probabilities we assume that every single node of the network can be the
origin of the walk with the same probability and  that the average runs over all sites $j=1,\dots,N$.  For the
individual return probability $\pi_{j,j}(t)$ we use the outer corner node $1$ as initial node.  As for the mean survival
probabilities, we will distinguish between two situations: (1) when there are three (four) trap nodes at the outer
corners of the gasket (carpet), see the red diamonds in Fig~\ref{fig:fractals}, and (2) when the three (four) trap nodes
are placed at the corners of the largest empty inner triangle (square) of the gasket (carpet), see the red squares in
Fig.~\ref{fig:fractals}. Since the quickest decay of $\Pi(t)$ for the linear networks studied in Ref.~\cite{Muelken2007}
is obtained when the trapping strength $\Gamma$ is  of the same order of magnitude as the coupling between the nodes, we
choose $\Gamma=1$ in all calculations involving traps.

\begin{figure}[ht!]
\includegraphics[width=\columnwidth]{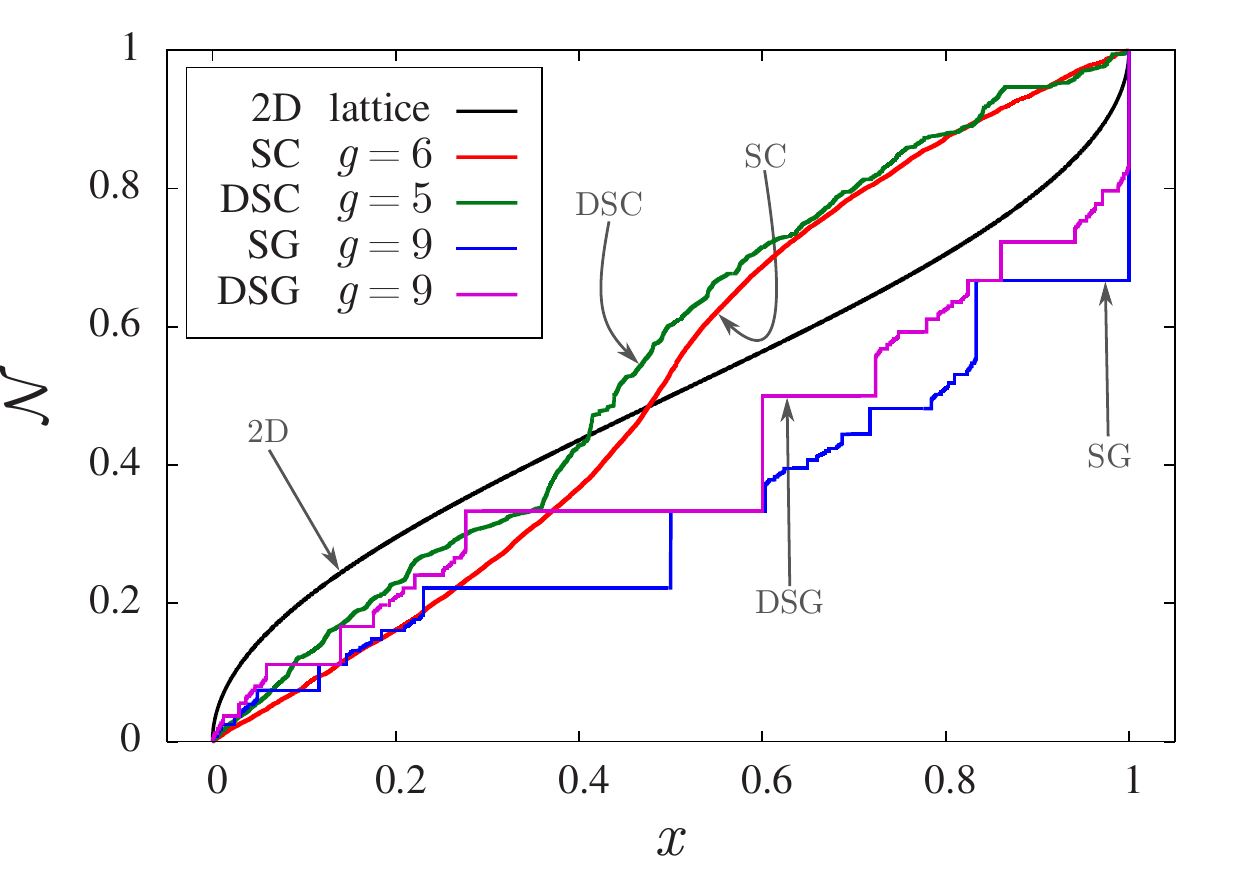}
\caption{(Color online) The eigenvalue counting function $\mathcal{N}(x)$, Eq.~(\ref{eq:ev_count}), for several systems
	under study, compared to  the simplest case of an infinite discrete square lattice, see text for details.}
\label{fig:Eigencount}
\end{figure}

Let us first consider systems without traps. Since the eigenvalue distributions are crucial for determining the global
efficiency measures, we start by considering the differences between our four fractal structures.  In Fig.
\ref{fig:Eigencount} we plot for several structures the normalized cumulative eigenvalue counting function
\begin{equation}
\mathcal{N}(x) = \frac{1}{N} \sum_{n=1}^{N} \theta(x - \frac{E_n}{E_{max}}),
\label{eq:ev_count}
\end{equation}
where $\theta(x)$ is the Heaviside-function. Now, $E_{min}=0$ is the smallest and $E_{max}$ the largest eigenvalue, hence,
 the range of $x$ is $[0,1]$.

Already here we can exemplify the role of highly degenerate eigenvalues. For large $N$ the eigenvalue counting function for an
$N\times N$ square lattice is a quite smooth function, which for $N \to \infty $ we plot as a reference in  Fig.~\ref{fig:Eigencount}.

Also the SC for $g=6$ leads to a quite smooth form for $\mathcal{N}(x)$. However,
$\mathcal{N}(x)$ for the DSC of $g=5$ displays marked steps, but its overall shape is close to the
one for the SC. For the SG for $g=9$ and its dual,  the DSG for $g=9$, $\mathcal{N}(x)$ has sharp discontinuities, which
reflect the presence of many highly degenerate eigenvalues. Already at this point we see a clear distinction between
gaskets and carpets: at similar $g$ the carpets do not have eigenvalues of such high degeneracy as the gaskets.

\section{Dual Sierpinski Gasket}\label{DSG}

We start by considering the DSG, see also Fig.~\ref{fig:fractals}(b). As the SG, the DSG is a deterministic fractal,
iteratively built up generation by generation. CTQW on DSG of different generations have been studied by us in
Ref.~\cite{Agliari2008}. We will recapitulate the major results, since we will use the DSG as a reference for our new
results presented below. In fact, DSG is special, in that its  eigenvalues, and hence its DOS can be determined
iteratively, in a simple way. This does not hold for the  other fractals considered here.

For DSG the results for the CTRW and CTQW return probabilities $p_{1,1}(t)$ and $\pi_{1,1}(t)$, along with the CTQW
lower bound $\big|\overline\alpha(t)\big|^2$ of $\overline{\pi}(t)$  (see Eqs.~(\ref{eq:ret-prob-cl}),  
(\ref{eq:ret-prob-q}), and (\ref{eq:lower_bound_q}), respectively) have been already presented in Ref. \cite{Agliari2008}.  There it
has been verified that for the classical average return probability, the decay to the equipartition value is determined
solely by $d_s$ \cite{alexander1982density}, having  namely $\overline p(t) \sim t^{-d_s/2}$. It follows that the
classical walk on DSG is recurrent and that the P\'olya number equals unity.  As we will show below for all the fractal
types considered here, such a quite simple law does not hold for CTQW.

Turning now to the quantum case and evaluating the lower bound $\big|\overline\alpha(t)\big|^2$ of $\overline{\pi}(t)$
of the quantum average return probability $\overline{\pi}(t)$, see Eq.~(\ref{eq:lower_bound_q}), it has been found in
\cite{Agliari2008} that its envelope does not show a strong dependence on the size of the DSG.  Since the two
eigenvalues $3$ and $5$ make up for about $1/3$ of all eigenvalues, they control most of the behavior of
$\overline\pi(t)$. Then $\rho(3)$ and $\rho(5)$ are known in closed form,
\begin{gather}
	\rho(3) = \frac{1}{2\cdot 3^g} \left ( 3^{g-1} + 3 \right )
\label{eq:DSGev-3}
\end{gather}
and
\begin{gather}	
	\rho(5) = \frac{1}{2\cdot 3^g} \left ( 3^{g-1} - 1 \right ) \, .
	\label{eq:DSGev-5}
\end{gather}
In particular,  also the long time average $\overline\chi_{lb}$ can be calculated exactly, 
based on Eq.~(\ref{eq:DOS-lta})
\begin{gather}
\label{eq:lta-dsg}
	\overline\chi_{lb} = \frac{1}{3^{2g}}\left[ 3^g\left(1+\frac{3^g}{14}\right)+\frac{10}{7}2^g -\frac{3}{2}\right] \, ,
\end{gather}
which for large $g$ is much larger than the equipartition value $3^{-g}$. 
The limit $g\to\infty$ yields
\begin{equation}
\lim_{g\to\infty} \overline\chi_{lb} = 1/14 \approx 0.0714 \, .
\end{equation}
For both highly degenerate eigenvalues, TABLE~\ref{table:Dual_gasket_eigenvalues} shows $\rho(3)$ and $\rho(5)$,
Eq.~(\ref{eq:intDOS}), for successive generations $g$ from $2$ to $8$, calculated  according to Eqs.~(\ref{eq:DSGev-3})
and (\ref{eq:DSGev-5}). Also the exact value of $\overline\chi_{lb}$, see Eq.~(\ref{eq:lta-dsg}), is shown. Both
$\rho(3)$ and $\rho(5)$ tend to the exact limiting value $1/6$, see Eqs.~(\ref{eq:DSGev-3}) and (\ref{eq:DSGev-5}),
rather fast, which, together with Eq.~(\ref{eq:DOS_localization}), means that the transport is quite inefficient.
\begin{table*}[ht!]
\begin{ruledtabular}
\begin{tabular}{crrc}
$g$ & $\rho(3)$ \hspace{4mm} & $\rho(5)$ \hspace{4mm} &
 $\overline\chi_{lb}$ \\
\hline
$2$ & $1/3 \approx 0.3333$ &  $1/9 \approx 0.1111$ & $0.2346$\\
$3$ & $2/9 \approx 0.2222$ & $4/27 \approx 0.1481$ & $0.1221$ \\
$4$ & $5/27 \approx 0.1852$ & $13/81 \approx 0.1605$ &  $0.0870$ \\
$5$ & $14/81 \approx 0.1728$ & $40/243 \approx 0.1646$ & $0.0763$ \\
$6$ & $41/243 \approx 0.1687$ & $121/729 \approx 0.1660$ & $0.0730$ \\
$7$ & $122/729 \approx 0.1674$ & $364/2187 \approx 0.1664$ & $0.0719$ \\
$8$ & $365/2187 \approx 0.1669$ & $1093/6561 \approx 0.1666$ & $0.0716$ \\
\end{tabular}
\caption{The $\rho(E)$ for the
	eigenvalues $E=3$ and $E=5$ and the long time average $\overline\chi_{lb}$ for different generations of the DSG.}
\label{table:Dual_gasket_eigenvalues}
\end{ruledtabular}
\end{table*}
\begin{table}[ht!]
\begin{ruledtabular}
\begin{tabular}{crr}
$g$ & $\Pi_\infty^{(1)}=N_0^{(1)}/N$ & $\Pi_\infty^{(2)}=N_0^{(2)}/N$ \\
\hline
2 & $1\,/\,9 \approx 0.111$ & 0 \\
3 & $9\,/\,27 \approx 0.333$ & $6\,/\,27 \approx 0.222$  \\
4 & $43\,/\,81 \approx 0.531$ & $36\,/\,81 \approx 0.444$ \\
5 & $165\,/\,243 \approx 0.679$ & $150\,/\,243 \approx 0.617$  \\
6 & $571\,/\,729 \approx 0.783$ & $540\,/\,729 \approx 0.741$ \\
7 & $1869\,/\,2187 \approx 0.855$ & $1806\,/\,2187 \approx 0.826$ \\
\end{tabular}
\caption{The asymptotic limit $\Pi_\infty$ of $\Pi(t)$ for DSG of  generations $g=2,3\dots,7$; case $(1)$: the traps are
	placed on the corner nodes, diamonds in Fig.~\ref{fig:fractals}(b); case $(2)$: the traps are placed on the central
	nodes, squares in Fig.~\ref{fig:fractals}(b), see text for details.}
\label{table:Dual_Gasket_survival}
\end{ruledtabular}
\end{table}

Now, we calculate for the DSG $\Pi_\infty^{(1)}$ and $\Pi_\infty^{(2)}$ using Eq.~(\ref{eq:pi-limit}).  In order to do
this, we numerically determine the eigenvalues of the non-Hermitian $\mathbf{H}_{\rm eff}$, paying particular  attention
to their imaginary parts $\gamma$. We do this using the MATLAB\,/\,GNU Octave eig() function, and in order to be
more precise, we employed the LAPACK zgeev() function in our Fortran code with quadruple precision. Despite these
efforts, the procedure may not be exact, however.  First, we cannot exclude the existence of very small, but nonzero
imaginary parts which are smaller than $10^{-31}$ and are set to zero. Second, numerical  errors may induce small
imaginary contributions where there should be none.  Thus, the values in our table for $\Pi_\infty$ may not be 
as exact as their form seems to imply.

Counting all the eigenvalues with vanishing imaginary part we then obtain $N_0$. From it we readily evaluate
$\Pi_\infty^{(1)}$ and $\Pi_\infty^{(2)}$, see Eq.~(\ref{eq:pi-limit}).   In the next sections, the same procedure will
be employed for the other fractals studied. The analysis of the data of TABLE \ref{table:Dual_Gasket_survival} shows
that  $\Pi_\infty^{(1)}$ and $\Pi_\infty^{(2)}$ increase with increasing $g$, which means that $N_0$ increases faster
than $N$.  Already for $g=7$, corresponding to a network of $N=2187$ nodes, the probabilities $\Pi_\infty^{(1)}$ and
$\Pi_\infty^{(2)}$ that the walker survives  within the network are close to $0.855$ and to $0.826$, respectively.  We
note that the values of $\Pi_\infty^{(2)}$ are somewhat below the ones for $\Pi_\infty^{(1)}$, implying that 
here traps on the periphery act somewhat less efficiently than centrally located traps.

\section{Sierpinski gasket}\label{SG}

While the DSG allows for partly analytical results, we have to resort to numerical calculations for the other structures
considered in this paper. We proceed our investigation with the SG. In analogy to our study on DSG, we start with the
CTRW and CTQW return probabilities as well as with the lower bound for the CTQW decay.  These quantities involve the
eigenvalues and (depending on the functions considered) sometimes also the eigenstates  of the Hermitian operators
$\mathbf{T}$ and $\mathbf{H}$, see Eqs.~(\ref{eq:ret-prob-cl})-(\ref{eq:ret-prob-q}). Unlike the DSG case, for SG
general  recursive expressions for the eigenvalues of $\mathbf{T}$ and $\mathbf{H}$ are not known. Therefore, we
calculate both the eigenvalues and the  eigenfunctions numerically.  For this, we again use the  MATLAB\,/\,GNU Octave
eig() and eigs() functions. For large generations,  we calculate only the spectrum in order to evaluate
$|\overline{\alpha}(t)|^2$ and the $D(E_m)$ degeneracy of the eigenvalue $E_m$ using  the filtered Lanczos algorithm in
C++ \cite{FILTLAN},  and the MATLAB\,/\,GNU Octave eigs() function.

\begin{figure}[ht!]
	\includegraphics[scale=0.7]{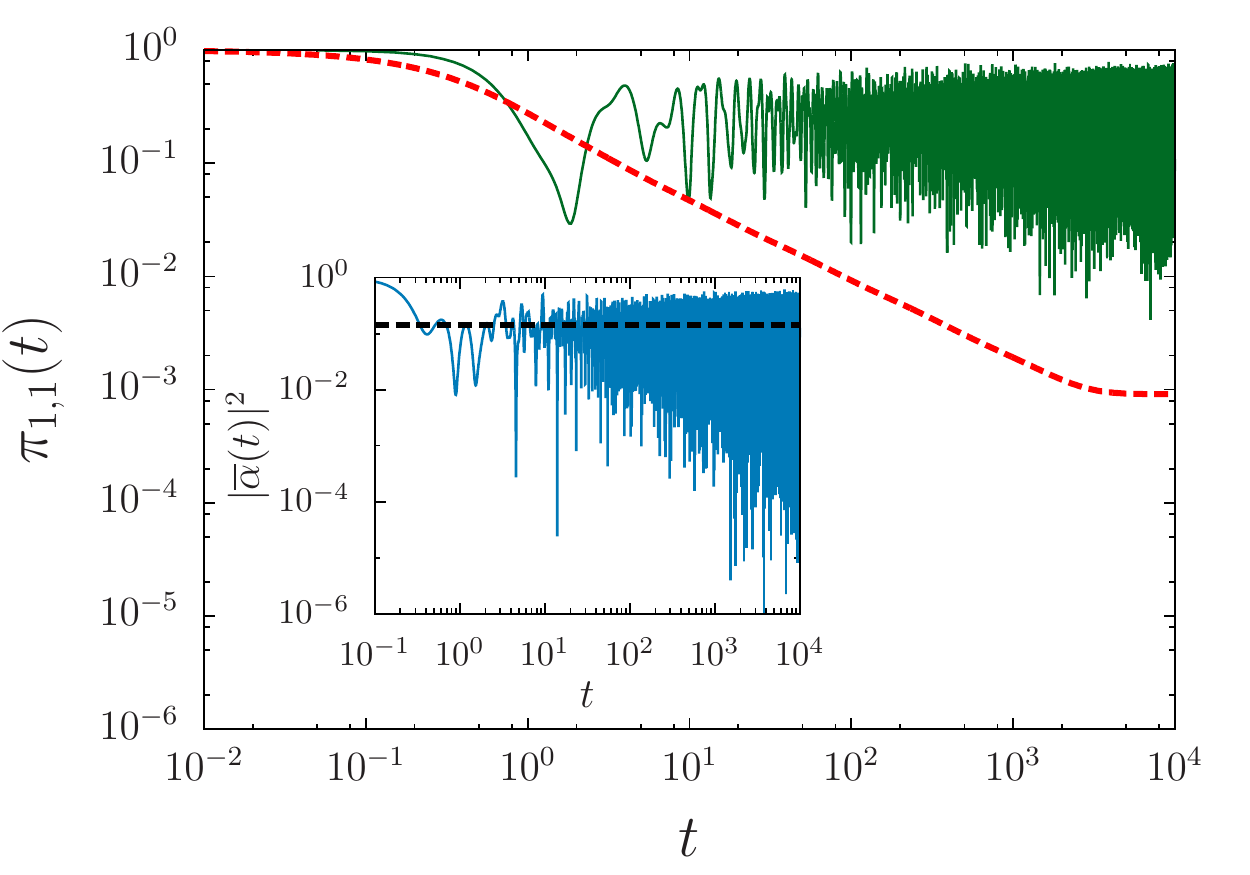}
\caption{(Color online) Quantum return probability $\pi_{1,1}(t)$ to the corner node $j=1$ (green solid line) along with its
	classical analogue (red dashed line) for the SG of $g=7$.  Inset: CTQW lower bound $\big|\overline\alpha(t)\big|^2$ of
	$\overline{\pi}(t)$ on the SG at $g=7$ (blue solid line) and $\overline\chi_{lb}$, the long time value (black dashed
	line).}
\label{fig:Gasket_ARP}
\end{figure}

Figure~\ref{fig:Gasket_ARP} shows the classical $p_{1,1}(t)$ and the quantum mechanical $\pi_{1,1}(t)$  return
probabilities to the initially excited node $j=1$ for a SG with $g=7$. The red dashed line in Fig.~\ref{fig:Gasket_ARP}
gives  the CTRW return probability $p_{1,1}(t)$.  While the algebraic decay of $p_{1,1}(t) \sim t^{-d_s/2}$ holds
asymptotically only for an infinite fractal, one can still recognize this  scaling behavior in an intermediate time
domain in Fig.~\ref{fig:Gasket_ARP},  before $p_{1,1}(t)$ saturates to the equipartition value $1/N$ at long times.
Figure \ref{fig:Gasket_ARP} also shows the exact quantum return probability $\pi_{1,1}(t)$ (green solid line). After an
initial decay to a local minimum, the return probability starts  to oscillate around its long time average. In the inset
of Fig.~\ref{fig:Gasket_ARP} we present the quantum mechanical lower bound $\big|\overline\alpha(t)\big|^2$ of the
quantum average return probability $\overline{\pi}(t)$; $\big|\overline\alpha(t)\big|^2$ does not decay eventually, but
shows strong oscillations with a long time average $\overline\chi_{lb}$ (dashed black line in Fig.~\ref{fig:Gasket_ARP})
which is orders of magnitude larger than $1/N$. Now, in contrast to CTRW, there is no apparent relation between the
spectral dimension and the return probability.
\begin{table}[ht!]
\begin{ruledtabular}
\begin{tabular}{crc}
$g$ &  $\rho(6)$ \hspace{5mm}  & $\overline\chi_{lb}$ \\
\hline
2 & 0 & 0.2778\\
3 & 3\,/\,15 = 0.2 & 0.1378\\
4 & 12\,/\,42 $\approx$ 0.2857 & 0.1179\\
5 & 39\,/\,123 $\approx$ 0.3171 & 0.1296\\
6 & 120\,/\,366 $\approx$ 0.3279  & 0.1374 \\
7 & 363\,/\,1095 $\approx$ 0.3315 & 0.1408\\
8 & 1092\,/\,3282 $\approx$ 0.3327 & 0.1421 \\
9 & 3279\,/\,9843 $\approx$ 0.3331 & 0.1426
\end{tabular}
\caption{The $\rho(E)$ for the
eigenvalue $E=6$ and the long time average $\overline\chi_{lb}$ for different
generations of the SG.}
\label{table:Gasket_eigenvalues}
\end{ruledtabular}
\end{table}
\begin{table}[ht!]
\begin{ruledtabular}
\begin{tabular}{crr}
$g$ &  $\Pi_\infty^{(1)}=N_0^{(1)}/N$ & $\Pi_\infty^{(2)}=N_0^{(2)}/N$  \\
\hline
2 & 0 & 0  \\
3 & $4\,/\,15 \approx 0.27$ & $1\,/\,15 \approx 0.067$\\
4 & $21\,/\,42 = 0.5$ & $15\,/\,42 \approx 0.357$\\
5 & $82\,/\,123 \approx 0.67$ & $70\,/\,123 \approx 0.569$ \\
6 & $285\,/\,366 \approx 0.78$ & $261\,/\,366 \approx 0.713$ \\
7 & $934\,/\,1095 \approx 0.85$ & $886\,/\,1095 \approx 0.809$\\
\end{tabular}
\caption{ The asymptotic limit $\Pi_\infty$ of $\Pi(t)$ for SG of generations $g=2,3\dots,7$; case $(1)$: the traps are
	placed on the corner nodes, diamonds in Fig.~\ref{fig:fractals}(a); case $(2)$: the traps are placed on the central
	nodes, squares in Fig.~\ref{fig:fractals}(a), see text for details.}
\label{table:Gasket_survival}
\end{ruledtabular}
\end{table}

The spectrum of the Hamiltonian already reveals whether CTQW shows localization. For different generations of the SG, we
calculate, based on Eq.~(\ref{eq:intDOS}),  the $\rho(E)$ of the highly degenerate eigenvalue $6$, $\rho(6)$ and, based
on the r.h.s. of  Eq.~(\ref{eq:DOS-lta}),  the long-time average  $\overline\chi_{lb}$. The data are presented in
TABLE \ref{table:Gasket_eigenvalues}. As in the case of the DSG, also for SG the $\rho(6)$ seem to converge with
increasing $g$ to the finite limiting value $1/3$. As before, such a relatively  large nonvanishing value lets us infer
that the transport is not very efficient.

We now turn to CTQW on SG in the presence of traps, process which introduces non-Hermitian operators. In TABLE
\ref{table:Gasket_survival}, we show $\Pi_{\infty}^{(1)}$ and $\Pi_{\infty}^{(2)}$ for two  situations, namely when the
traps are placed on the corners and when the traps are placed in the center of the structure, see
Fig.~\ref{fig:fractals}(a) for details.  TABLE \ref{table:Gasket_survival} suggests that the situation is quite similar
to the one for the DSG: the higher $g$ the less it is probable that the excitation will be absorbed even after a
very long time, see the increase in the $\Pi_\infty$-values. However, the amount of excitation which stays localized in
the network is higher in case $(1)$ than in case $(2)$, since  $\Pi_\infty^{(1)}>\Pi_\infty^{(2)}$.

\section{Dual Sierpinski carpet}\label{DSC}

 We continue our analysis by considering the SC and the DSC. We start with the DSC, the spectrum and harmonic functions
of which have been considered recently \cite{Carpet_homotopies,Carpet_eigenvalues}.

\begin{figure}[ht!]
	\includegraphics[scale=0.7]{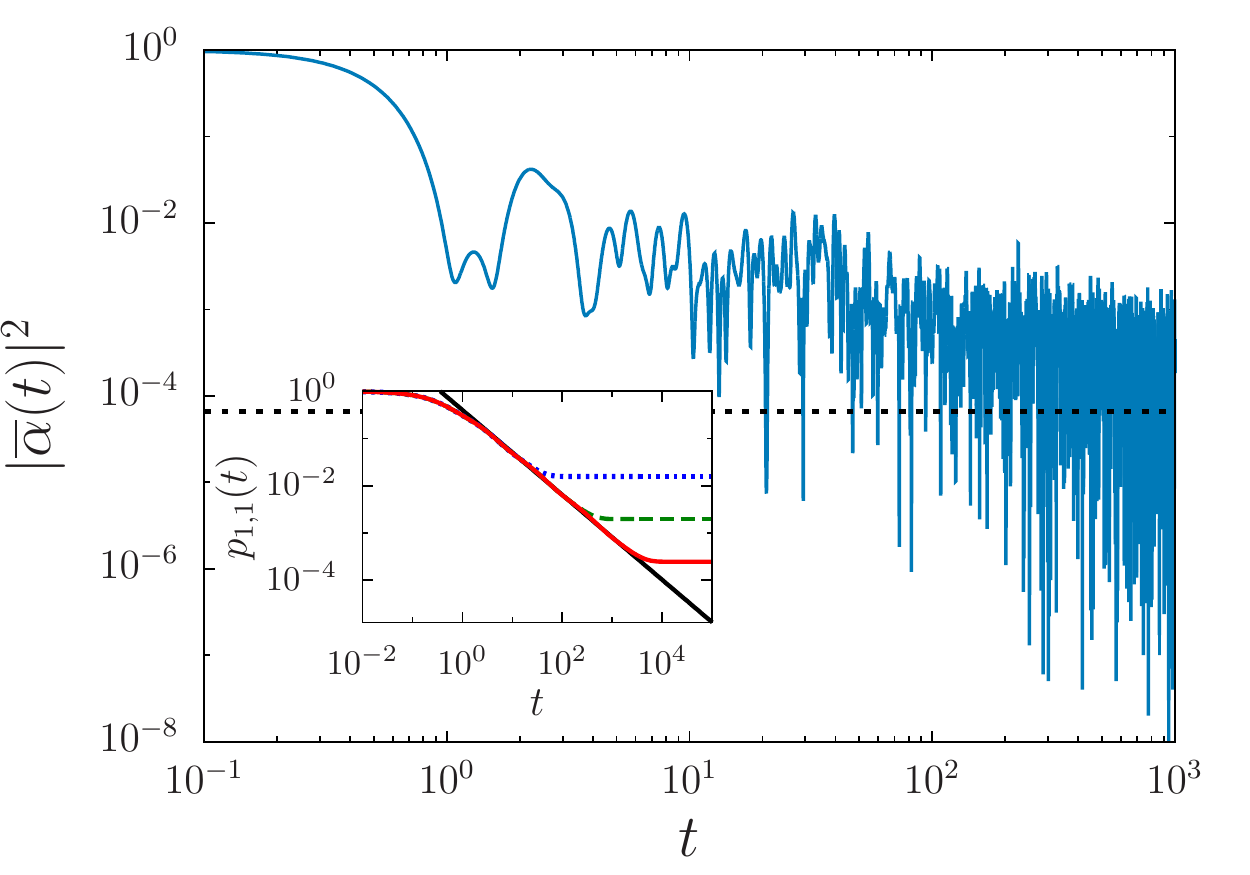}
	\caption{(Color online) The average return amplitude $|\overline\alpha(t)|^2$ on the DSC of $g=5$ (blue solid line) and the 
	long time average (black dotted line). Inset:
	classical return probability $p_{1,1}(t)$ for DSC of $g=2, 3$, and $4$ (dotted blue, dashed green, solid red line, 
respectively) as well as the fitted decay (solid straight black line), $0.4\cdot t^{-1.8/2}$.}
\label{fig:Dual_carpet_longtime}
\end{figure}

Figure~\ref{fig:Dual_carpet_longtime} presents the lower bound $|\overline\alpha(t)|^2$ of the quantum
$\overline{\pi}(t)$ for $g=5$, see Eq.~(\ref{eq:lower_bound_q}). The inset of  Figure~\ref{fig:Dual_carpet_longtime}
depicts the classical return probability $p_{1,1}(t)$ given by Eq.~(\ref{eq:ret-prob-cl}). We note that at intermediate
times  $p_{1,1}(t)$ shows an algebraic decay with slope $d_s/2$. Furthermore, $|\overline\alpha(t)|^2$ displays at short
to intermediate time a decay of the maxima, while at longer times it slowly approaches $\overline\chi_{lb}$, given in
Fig.~\ref{fig:Dual_carpet_longtime} through a dotted line around which it oscillates.   Given that for DSC
$\overline\chi_{lb}$ is much smaller than the corresponding $\overline\chi_{lb}$ for SG and for DSG, we infer that
localization effects are smaller for DSC than for SG and for DSG.
\begin{table}[ht!]
\begin{ruledtabular}
\begin{tabular}{crc}
$g$ &  $\rho(3)$ \hspace{10mm} & $\overline\chi_{lb}$ \\
\hline
2 & $2\,/\,64 \approx 3.13\times 10^{-2}$ & $2.44\times 10^{-2}$\\
3 & $4\,/\,512 \approx 7.81\times 10^{-3}$ & $2.98\times 10^{-3}$\\
4 & $20\,/\,4096 \approx 4.88\times 10^{-3}$ & $3.89\times 10^{-4}$\\
5 & $148\,/\,32768 \approx 4.52\times 10^{-3}$ & $6.60\times 10^{-5}$\\
6 & $1172\,/\,262144 \approx 4.47\times 10^{-3}$ &  -- 
\end{tabular}
\caption{The $\rho(3)$ for the
eigenvalue $E=3$ and the long time average $\overline\chi_{lb}$ for different
generations of the DSC.}
\label{table:Dual_carpet_eigenvalues}
\end{ruledtabular}
\end{table}
\begin{table}[ht!]
\begin{ruledtabular}
\begin{tabular}{crr}
$g$ &  $\Pi_\infty^{(1)}=N_0^{(1)}/N$ & $\Pi_\infty^{(2)}=N_0^{(2)}/N$ \\
\hline
2 & $15\,/\,64 \approx 0.234$ & $ 14\,/\,64 \approx 0.219$ \\
3 & $126\,/\,512 \approx 0.246$ & $126\,/\,512 \approx 0.246$ \\
4 & $1030\,/\,4096 \approx 0.251$ & $1030\,/\,4096 \approx 0.251$  \\
\end{tabular}
\caption{The asymptotic limit $\Pi_\infty$ of $\Pi(t)$ for DSC of  generations $g=2,3$, and $4$; case $(1)$: the traps
	are placed on the corner nodes, diamonds in Fig.~\ref{fig:fractals}(d); case $(2)$: the traps are placed on the central
	nodes, squares in Fig.~\ref{fig:fractals}(d), see text for details.}
\label{table:dual_carpet_survival}
\end{ruledtabular}
\end{table}

However, from the above results we cannot deduce whether the walk is recurrent or not. Therefore, for DSC we again
consider the spectrum of $\mathbf{T}$ and $\mathbf{H}$ and calculate, based on Eq.~(\ref{eq:intDOS}), $\rho(E)$  for the
most highly degenerate eigenvalue, see TABLE~\ref{table:Dual_carpet_eigenvalues}. Clearly, our calculations are limited
by computational power and for the DSC we could not obtain results for $g$ larger than $6$; for $g=6$ there are
already $N=262144$ nodes in the network. It seems as if for very large $g$ the $\rho(3)$ series will converge to a value
somewhat above $4.4\times10^{-3}$. This finite limit  again seems to indicate  that there is localization in the system,
so that CTQW may be recurrent in general.

Now, let us consider the CTQW trapping process for the DSC. TABLE \ref{table:dual_carpet_survival} presents
$\Pi_{\infty}$ for two different trap placements  on DSC of $g=2,3$, and $4$, as shown in
Fig.~\ref{fig:fractals}(d).  Our calculations of the mean survival probabilities $\Pi(t)$ and their asymptotic values
$\Pi_{\infty}$ do not allow for a clear-cut statement for the DSC.  The first thing to note is that $\Pi_{\infty}^{(1)}$
and $\Pi_{\infty}^{(2)}$ are very similar and that with increasing $g$  their values stay rather constant. This again is
only a weak indication that also the DSC shows localization.

\section{Sierpinski carpet}\label{SC}
 
\begin{figure}[ht!]
	\includegraphics[scale=0.7]{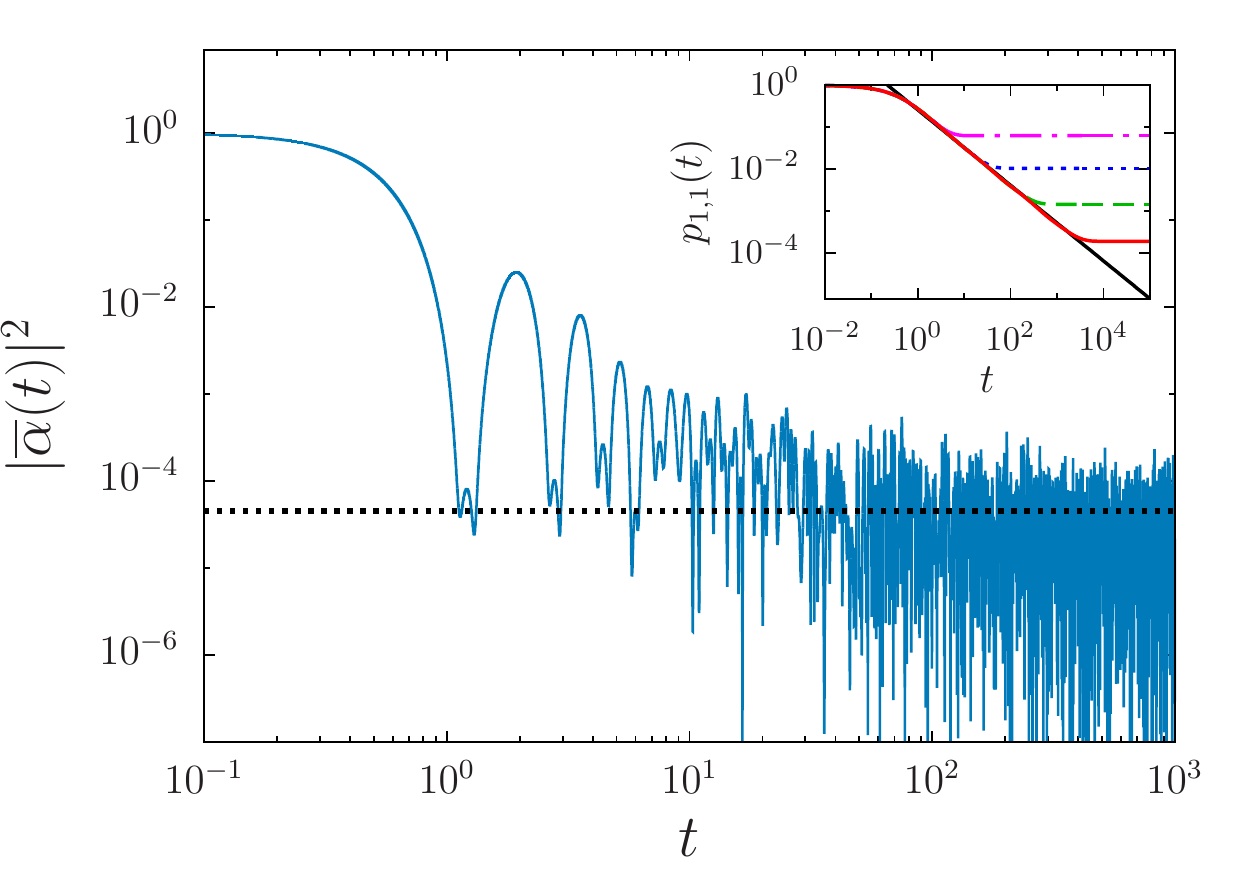}
	\caption{(Color online) Average return amplitude on the SC at $g=6$ (blue solid line) and the long time average 
	$\overline\chi_{lb}$ (black dotted line). Inset: classical return probability to the corner node for the SC at $g=2$, $3$, $4$, 
	and $5$ (pink dash dotted, blue dotted, green dashed, red solid line, respectively) and the decay according to the spectral 
	dimension $d_s$ (black straight solid line).}
\label{fig:Carpet_average}
\end{figure}

Let us now consider the transport properties of CTRW and of CTQW on SC. We start again by calculating the lower bound
$|\overline\alpha(t)|^2$ of the quantum average return probability  $\overline{\pi}(t)$, Eq.~(\ref{eq:lower_bound_q}),
see Fig.~\ref{fig:Carpet_average} for the DSC with $g=6$. The inset shows the behavior of the corresponding CTRW
$p_{1,1}(t)$ for various $g$,  which for intermediate times scales with $d_s$ as expected.
\begin{table}[ht!]
\begin{ruledtabular}
\begin{tabular}{crc}
$g$ &  $\rho(4)$  \hspace{10mm} & $\overline\chi_{lb}$\\
\hline
$2$ & $3\,/\,16 \approx 1.88\times 10^{-2}$ & $1.25\times 10^{-1}$ \\
$3$ & $6\,/\,96 = 6.25\times 10^{-2}$  & $1.89\times 10^{-2}$ \\
$4$ & $8\,/\,688 \approx 1.16\times 10^{-2}$ & $2.29\times 10^{-3}$ \\
$5$ & $16\,/\,5280 \approx 3.03\times 10^{-3}$ & $2.92\times 10^{-4}$ \\
$6$ & $128\,/\,41584 \approx 3.08\times 10^{-3}$ & $4.54\times 10^{-5}$
\end{tabular}
\caption{The $\rho(E)$ for the
eigenvalue $E=4$ and the long time average $\overline\chi_{lb}$ for different
generations of the SC.}
\label{table:Carpet_eigenvalues}
\end{ruledtabular}
\end{table}
\begin{table}[ht!]
\begin{ruledtabular}
\begin{tabular}{crr}
$g$ &  $\Pi_\infty^{(1)}=N_0^{(1)}/N$ & $\Pi_\infty^{(2)}=N_0^{(2)}/N$  \\
\hline
$2$ & $2\,/\,16  = 0.125$ & $2\,/\,16=0.125$ \\
$3$ & $23\,/\,96 \approx 0.240$ & $22\,/\,96 \approx 0.230$\\
$4$ & $168\,/\,688 \approx 0.244$ & $168\,/\,688 \approx 0.244$\\
$5$ & $1314\,/\,5280 \approx 0.249$ & $1314\,/\,5280 \approx 0.249$ \\
\end{tabular}
    \caption{The asymptotic limit $\Pi_\infty$ of $\Pi(t)$ for SC of generations $g=2,3, 4$, and $5$; case $(1)$: the
		traps are placed on the corner nodes, diamonds in Fig.~\ref{fig:fractals}(c); case $(2)$: the traps are placed on the
		central nodes, squares in Fig.~\ref{fig:fractals}(c), see text for details.}
\label{table:carpet_survival}
\end{ruledtabular}
\end{table}

Again the finite size of the network does not allow for a definite statement about the recurrent behavior of the CTQW.
Therefore, we again calculate $\rho(E)$, see Eq.~(\ref{eq:intDOS}), for the highly degenerate eigenvalue $E=4$; the
corresponding values are displayed in TABLE \ref{table:Carpet_eigenvalues}.  There we also show the long time average
$\overline\chi_{lb}$ calculated using the r.h.s. of Eq.~(\ref{eq:DOS-lta}). Similar to the DSC case, there is no strong
evidence that CTQW on SC show localization.  Again, as for DSC, the limiting value of $\rho(4)$ lies above
$3\times10^{-3}$,  such that at this point we cannot make any precise statement.  We could not obtain results for larger
generations because (at present) we do not have the computational facilities to calculate $\rho(4)$ for $g>7$; the size
of the corresponding matrix for a DSC of $g=7$ is already larger than $300\,000\times 300\,000$.

Considering now absorption processes, when there are traps placed on three nodes of each network (see, e.g.,
Fig.~\ref{fig:fractals}(c)),  we again calculate the quantum mechanical limit $\Pi_\infty$ of $\Pi(t)$; the
corresponding results are given in TABLE~\ref{table:carpet_survival}.  Here we find that CTQW on SC are quite different
from those on the gaskets, but that they are similar to CTQW on DSC: both, $\Pi_\infty^{(1)}$ and $\Pi_\infty^{(2)}$,
show a slow increase with increasing $g$.

\section{Summary}\label{summary}	

Our  analysis of CTQW over different types of Sierpinski fractals revealed interesting aspects of quantum mechanical
transport. At first, for SG and for DSG we find strong localization
effects, supported by the fact that the long time averages $\overline\chi_{lb}$ approach a finite limiting value  with increasing
$g$, see TABLES~\ref{table:Dual_gasket_eigenvalues} and \ref{table:Gasket_eigenvalues}. For the carpets we cannot make a
definite statement based on our numerical results for $\overline\chi_{lb}$ for generations up to $g=6$. 

Turning now to the DOS and monitoring in each case the eigenvalues with the highest degeneracy, we find that
$\rho(3)$ and $\rho(5)$ for DSG and $\rho(6)$ for SG  tend with growing $g$ each to a constant,  quite significant
value, see  TABLES~\ref{table:Dual_gasket_eigenvalues} and \ref{table:Gasket_eigenvalues}. This supports our
view that the walkers are localized  even for very large $g$, in line with Refs.~\cite{mulken2011continuous,
Muelken2006}. For the carpets, for which the eigenvalue with the highest degeneracy is $3$ for DSC  and $4$ for SC, we
find that the corresponding values $\rho(3)$ for the DSC and $\rho(4)$ for the SC are rather small, which renders a
clear cut decision on localization difficult. We hence conclude that one needs much larger carpets than the ones we
could (at present) numerically handle, in order to attain a definite conclusion.
\begin{center}
\begin{figure*}[ht!]
	\centerline{\includegraphics[width=0.33\linewidth]{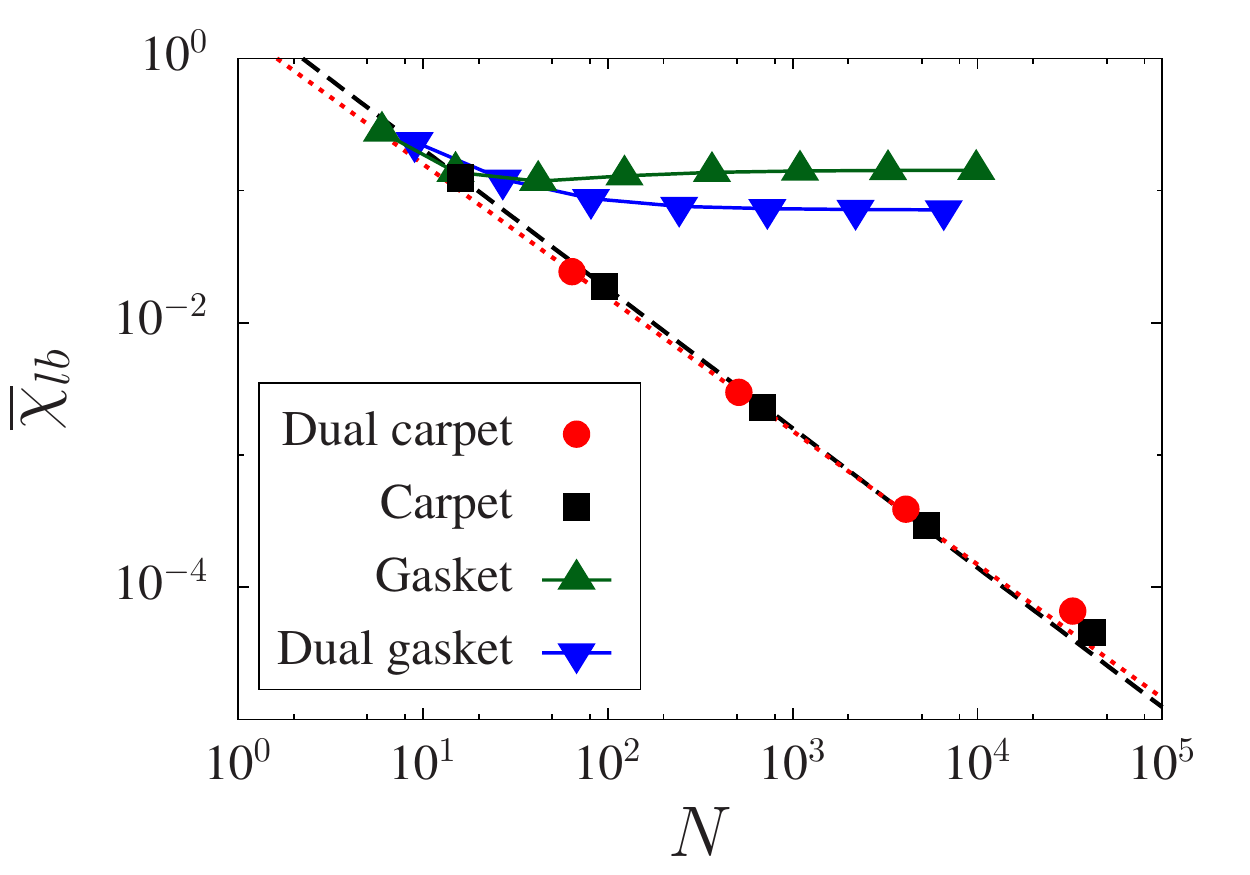}
	\includegraphics[width=0.33\linewidth]{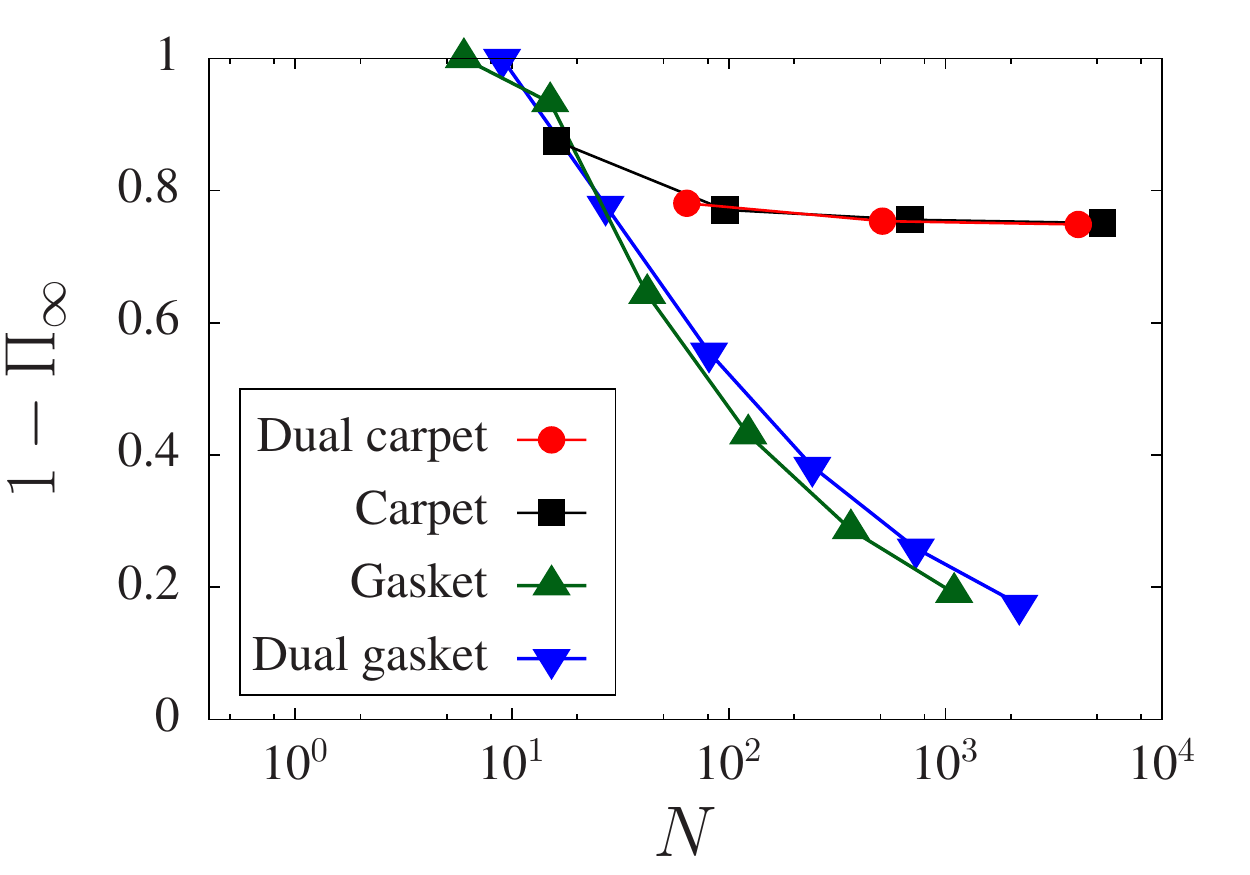}
	\includegraphics[width=0.33\linewidth]{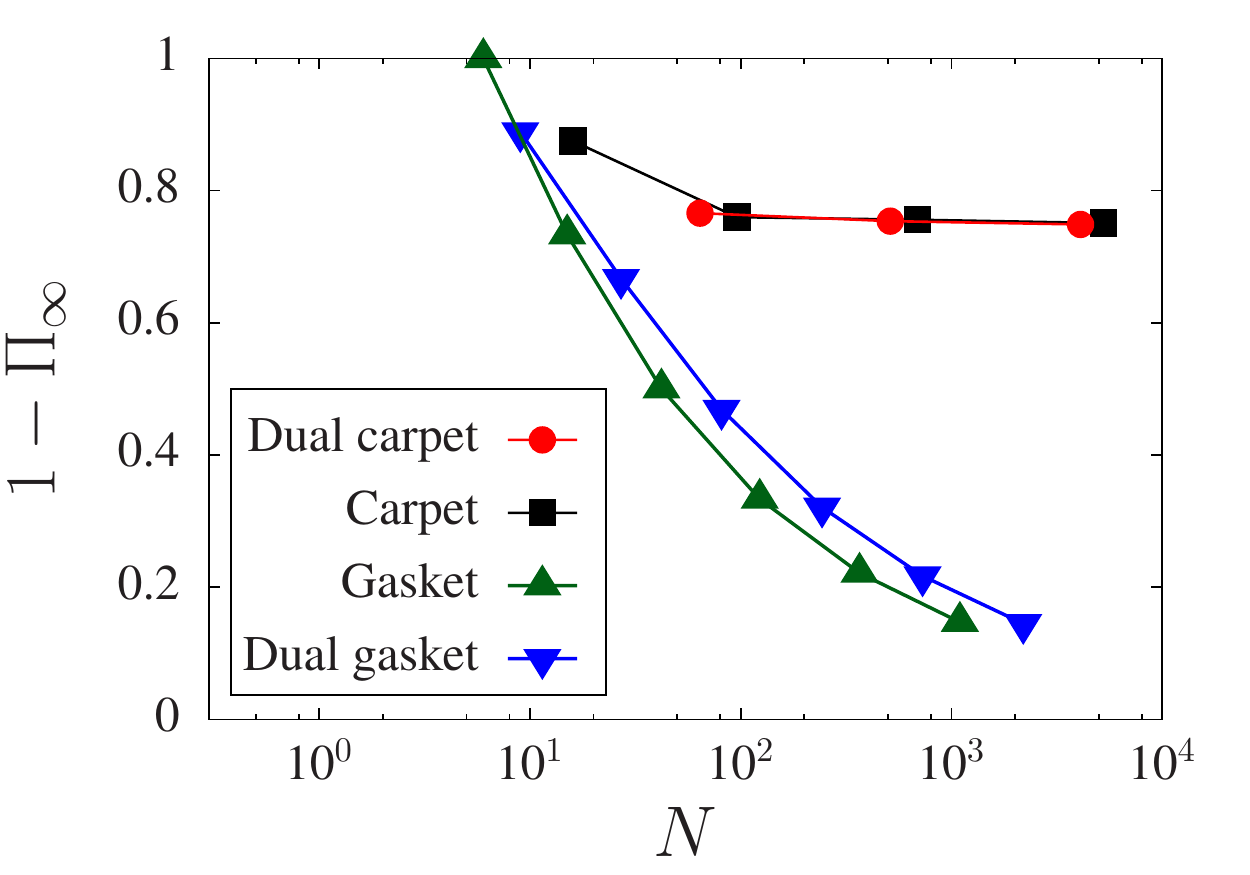}}
\caption{(Color online)  Long time average $\overline\chi_{lb}$ of $\pi_{1,1}(t)$ and probability
 $1-\Pi_{\infty}$ for an excitation to get trapped as a function of the number of nodes.}
\label{fig:Conclusion}
\end{figure*}
\end{center}

These results are confirmed by our findings for the mean survival probabilities $\Pi^{(j)}_\infty$ ($j=1,2$) for both
arrangements of traps. Here, the mean survival probability for SG and DSG increases with $N$, meaning that for larger
networks it becomes less and less probable that the excitation will leave the network. Since in each case we consider
only three trap nodes, with increasing $g$ the number of nodes without traps increases. Hence, due to localization, an
excitation starting from a node far away from the traps will not be able to reach them. For SC and DSC, the mean
survival probabilities show only a slight increase  with $g$ (for the network sizes considered here), thus the
probability of being trapped is almost independent of the size of the network, see
TABLES~\ref{table:dual_carpet_survival} and \ref{table:carpet_survival}.
In this respect, it would be also interesting to investigate the effect when the number of traps also increases
with $g$.

In addition and in contrast to the corresponding classical CTRW, for CTQW there is no apparent scaling behavior with the
spectral dimension $d_s$. This is obvious for the gaskets, see Ref.~\cite{Agliari2008} for DSG and
Fig.~\ref{fig:Gasket_ARP} for SG. However, for the carpets one might still argue that for large generations $g$ such a
scaling could exist for the envelope of $|\overline\alpha(t)|^2$ at intermediate times, see
Figs.~\ref{fig:Dual_carpet_longtime} and \ref{fig:Carpet_average}. But from our numerical results for generations up to
$g=7$ we cannot draw this conclusion.

Nevertheless, the long time behavior of the CTQW with and without traps reveals clear-cut differences between gaskets
and carpets for the generations studied here. While classically the difference is only manifested in a different scaling
according to $d_s$, quantum-mechanically we find there appears to be a fundamental
difference between gaskets and carpets - at least for the finite networks studied here. Our main results are summarized
in Fig.~\ref{fig:Conclusion}, where we plot, as a function of the number of nodes $N$, the long time average of the
lower bound of the averaged return probability, $\overline\chi_{lb}$, and (for practical purposes) the long time value
of the mean trapping probability, i.e., $1-\Pi^{(j)}_\infty$ ($j=1,2$), for the two situations of trap arrangements.

Already the spectra of the gaskets and of the carpets are significantly different and contain - in principle - all the
essential information. Now, for  the classical CTRW only the low-energy part of the spectrum is important for the
intermediate-to-long time behavior, whereas for CTQW the whole spectrum matters. Clearly, further investigations of these facts 
are in order, but go beyond the scope of the present paper.  In general, a systematic study of the importance of the so-called 
ramification number  (the number of nodes/bonds which has to be removed in order for the fractal to fall apart), 
along with a careful analysis of  localized eigenstates (e.g., ``dark states'' for the trap) deserves further studies.
From the point of view of bond percolation, we have studied localized eigenstates of two dimensional lattices with traps and 
with randomly placed bonds \cite{Anishchenko_Geometrical_2013}. There, we found that the localization feature is also
mirrored in the survival probablity.
While these aspects have been already touched upon  in Ref.~\cite{domany1983solutions} for the SG, no detailed analysis
has been carried out for the SC.

\section*{Acknowledgments}

We thank Dr.\ J\'anos Asb\'oth for stimulating discussions. We further thank the Deutscher Akademischer Austauschdienst
(DAAD Grant No. 56266206 and project no.\ 40018) for supporting mutual visits of the Freiburg and the Budapest groups.
Z.D. and T.K. acknowledge support by the Hungarian Scientific Research Fund (OTKA) under contract nos.\ K83858 and
NN109651, and the Hungarian Academy of Sciences (Lend\"ulet Program, LP2011-016). A.A., A.B., and O.M. acknowledge
support from the Deutsche Forschungsgemeinschaft (DFG Grant No. MU2925/1-1), from  the Fonds der Chemischen Industrie,
and from the Marie Curie International Research Staff Exchange Science Fellowship within the 7th European Community
Framework Program SPIDER (Grant No. PIRSES-GA-2011-295302).

\bibliography{references} 

\end{document}